\documentclass[10pt]{article}
\usepackage[T1]{fontenc}
\usepackage[utf8]{inputenc}
\usepackage[margin=1in]{geometry}
\usepackage{graphicx}
\usepackage{amsmath}
\usepackage{booktabs}
\usepackage[hidelinks]{hyperref}
\usepackage[skip=4.5pt]{parskip}
\usepackage{microtype}
\usepackage{array}

\title{\bfseries Ask a Language Model for Lottery Numbers:\\
Concentration in Repeated Six-of-49 Outputs}
\author{Dmitrij \.Zatuchin\\[2pt]
\small Department of Information Technologies, Estonian Entrepreneurship University\\[-2pt]
\small of Applied Sciences (EUAS), Tallinn, Estonia; and Rankfor.AI, Tallinn, Estonia\\[-2pt]
\small \texttt{dmitrij.zatuchin@eek.ee}}
\date{September 2026}

\begin{document}
\maketitle

\begin{abstract}
\noindent We evaluate six language-model configurations on requests for six distinct random
integers from 1-49. Across 1{,}200 attempted calls using four English prompt variants,
1{,}184 responses yielded valid tickets. Effective diversity of number frequencies ranged
from 9.9 to 18.0, compared with simulated fifth-percentile thresholds of 46.6-46.7 under
independent uniform six-of-49 sampling at the corresponding sample sizes. Systems produced
8-93 distinct unordered tickets, and their modal tickets accounted for 22.5-68.0\% of valid
responses. Two archived Polish Lotto samples provided a physical-lottery comparison, with
effective diversities of 41.1 and 41.4 at smaller sample sizes. These results demonstrate
substantial concentration under the tested deployment settings. They do not identify its
mechanism or establish performance under other prompts, temperatures, or tool
configurations.
\end{abstract}

\section{Design and benchmarks}

Sampling temperature makes a language model's output vary from run to run, and that
variation is easy to read as randomness. We evaluate outputs against independent uniform
sampling of six distinct integers from 1-49. Archived Polish Lotto results provide a
physical-lottery comparison: the operator's published draw statistics for 1 June to
31 August 1999 (26 draws, 156 numbers) and the same window in 2026 (39 draws, 234 numbers),
two same-season samples 27 years apart.

Six engines (gpt-5.6-luna, claude-sonnet-5, gemini-3.7-flash, grok-4.5, mistral-large,
Perplexity sonar) each received 200 attempted calls in English, distributed evenly across
four prompt variants, 50 calls per variant; the exact prompt texts are in
Appendix~\ref{app:prompts}. Every call opened a fresh context. We evaluated the
provider-default configurations used by the collector; this experiment did not vary
decoding settings systematically. sonar performs retrieval by default; the other five
configurations received no search parameters. Calls ran in randomised order on
1 September 2026, and the per-engine deployment record is in Appendix~\ref{app:deploy}.

Every source is scored by the effective diversity of number frequencies,
$N_{\mathrm{eff}} = 1/\sum_{i=1}^{49} s_i^2$, where $s_i$ is the share of all returned
numbers landing on $i$. A source using all 49 numbers equally scores 49; a source always
returning the same six scores 6. Finite samples score below 49 even under uniform
sampling: the simulated median is 38.5 at 26 tickets and 41.5 at 39, so the archived
samples' values near 41 reflect sample size, never a 41-number ceiling.

The test is Monte Carlo against the uniform ticket sampler. For each source we simulate
$B = 20{,}000$ datasets of independent uniform tickets, six distinct numbers per ticket,
at that source's valid-ticket count, and compute
$p_{\mathrm{MC}} = (1 + \#\{\mathrm{sim}\ N_{\mathrm{eff}} \le \mathrm{observed}\})/(B+1)$,
one-sided with ties counted. When no simulated dataset is as concentrated we report
$1/20{,}001 \approx 0.00005$, never zero. Seeds are fixed per sample
size (20260904 to 20260908) and the script \texttt{compute\_revision.py} distributed with
this note reproduces every value. An OR rule combining $N_{\mathrm{eff}}$ with a
top-number count has a false-positive rate between 6.1 and 7.7 percent at these sample
sizes, above the nominal 5, so $N_{\mathrm{eff}}$ alone is the test and the top-number
count stays a descriptive diagnostic (Table~\ref{tab:neff}).

\begin{figure}[p]
\centering
\includegraphics[width=0.96\textwidth]{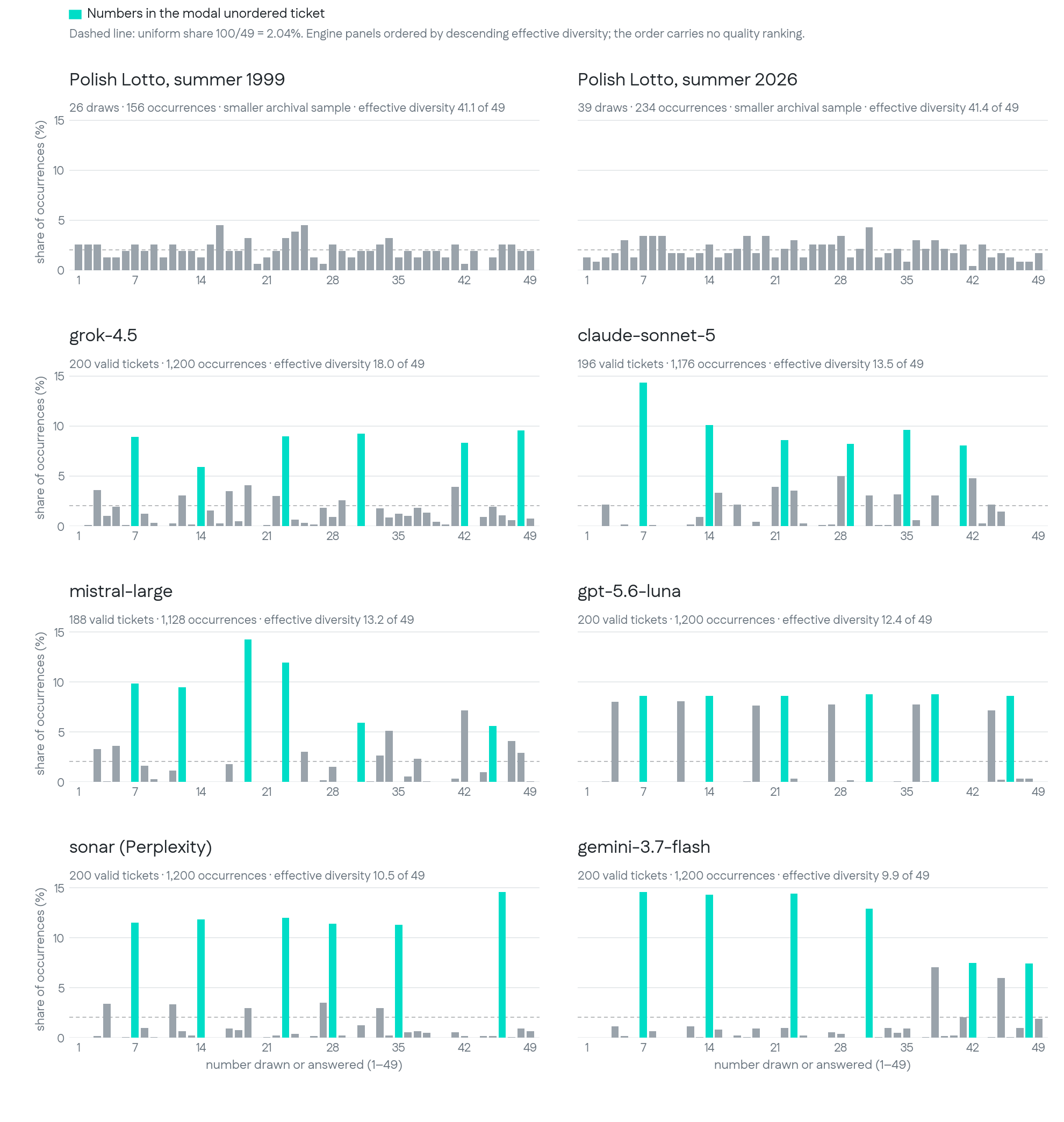}
\caption{Share of number occurrences, 1 to 49, on a shared y-scale. Top row: the two
archived Polish Lotto samples, 26 and 39 draws, marked as smaller archival samples. Rows
two to four: six engines, 188 to 200 valid tickets each, four English prompt variants,
provider-default settings, fresh context per call, 1 September 2026. Mint bars mark the
numbers in each engine's modal unordered ticket. The dashed line is the uniform share
$100/49 = 2.04\%$. Engine panels are ordered by descending effective diversity; the order
is presentational and carries no quality ranking.}\label{fig:lotto}
\end{figure}

\clearpage

\section{Results}

\textbf{Spread of numbers.} Figure~\ref{fig:lotto} carries the distributions and
Table~\ref{tab:neff} the test. All six engines score $N_{\mathrm{eff}}$ between 9.9 and
18.0 against simulated fifth-percentile floors of 46.6 to 46.7 at their sample sizes, with
$p_{\mathrm{MC}} = 0.00005$ each. The two archived samples score 41.1 and 41.4 at 26 and
39 draws, with $p_{\mathrm{MC}} = 0.95$ and 0.49, compatible with the uniform benchmark at
their sample sizes.

\begin{table}[t]
\centering
\caption{Effective diversity of number frequencies per source, against the uniform
ticket-sampler null: $B = 20{,}000$ simulated datasets of independent uniform
six-distinct-of-49 tickets at each source's valid-ticket count, seeds 20260904 to
20260908 by sample size. The floor is the simulated 5th percentile of $N_{\mathrm{eff}}$;
$p_{\mathrm{MC}} = (1+\#\{\mathrm{sim} \le \mathrm{obs}\})/(B+1)$, one-sided, reported as
$1/20{,}001 = 0.00005$ when no simulated dataset was as concentrated. The top-number count
is a descriptive diagnostic, shown with its simulated 95th percentile.}\label{tab:neff}
\small
\begin{tabular}{lrrrrrr}
\toprule
source & tickets & numbers & $N_{\mathrm{eff}}$ of 49 & floor & $p_{\mathrm{MC}}$ & top count (p95) \\
\midrule
Polish Lotto, summer 1999 & 26 & 156 & 41.1 & 35.7 & 0.95 & 7 (9) \\
Polish Lotto, summer 2026 & 39 & 234 & 41.4 & 39.2 & 0.49 & 10 (12) \\
grok-4.5 & 200 & 1{,}200 & 18.0 & 46.7 & 0.00005 & 115 (40) \\
claude-sonnet-5 & 196 & 1{,}176 & 13.5 & 46.7 & 0.00005 & 169 (39) \\
mistral-large & 188 & 1{,}128 & 13.2 & 46.6 & 0.00005 & 161 (38) \\
gpt-5.6-luna & 200 & 1{,}200 & 12.4 & 46.7 & 0.00005 & 105 (40) \\
sonar (Perplexity) & 200 & 1{,}200 & 10.5 & 46.7 & 0.00005 & 175 (40) \\
gemini-3.7-flash & 200 & 1{,}200 & 9.9 & 46.7 & 0.00005 & 175 (40) \\
\bottomrule
\end{tabular}
\end{table}

\textbf{Tickets.} The stronger result sits at the level of whole answers
(Table~\ref{tab:tickets}). Tickets are canonicalised as sorted six-number sets, so two
answers with the same six numbers in different orders count as one ticket. There are
13{,}983{,}816 such tickets, and under independent uniform sampling the probability of any
repeated ticket in 200 draws is 0.0014, about one in 703. gpt-5.6-luna produced 8 distinct
tickets in 200 valid answers and returned its most frequent one, 7~14~22~31~38~46, in 101
of them. sonar returned the same six numbers in 68.0\% of its answers. The smallest number
in every system's modal unordered ticket was 7. gemini-3.7-flash and grok-4.5 share the
identical modal ticket, 7~14~23~31~42~48, a descriptive observation with no mechanism
implied.

\begin{table}[t]
\centering
\caption{Whole-ticket behaviour per engine. Valid answers are the parsed responses of 200
attempted calls, pooled over the four prompt variants; the modal share is over valid
answers.}\label{tab:tickets}
\small
\begin{tabular}{lrrlr}
\toprule
engine & valid answers & distinct tickets & modal ticket & modal share \\
\midrule
gpt-5.6-luna & 200 & 8 & 7 14 22 31 38 46 & 50.5\% \\
mistral-large & 188 & 24 & 7 12 19 23 31 45 & 26.6\% \\
sonar (Perplexity) & 200 & 30 & 7 14 23 28 35 46 & 68.0\% \\
claude-sonnet-5 & 196 & 36 & 7 14 22 29 35 41 & 26.0\% \\
gemini-3.7-flash & 200 & 36 & 7 14 23 31 42 48 & 38.0\% \\
grok-4.5 & 200 & 93 & 7 14 23 31 42 48 & 22.5\% \\
\bottomrule
\end{tabular}
\end{table}

\textbf{Prompt variants.} Pooled results are results for the specified mixture of four
prompts. The per-variant table in Appendix~\ref{app:prompts} shows concentration in every
one of the 24 cells and a wide spread across variants: gpt-5.6-luna returned 2 to 3
distinct tickets with a 96\% modal share under the base and lottery phrasings, and
grok-4.5 moved from an 8\% modal share ($N_{\mathrm{eff}}$ 23.1) under the base phrasing
to 72\% ($N_{\mathrm{eff}}$ 7.4) when the word lottery appeared in the prompt.

\textbf{Ordering.} Most systems returned numbers in ascending order. We treat ordering as
a formatting characteristic and analyse ticket identity as an unordered six-number set.
gemini-3.7-flash and gpt-5.6-luna returned the six numbers already sorted in 100\% of
parsed answers, claude-sonnet-5 in 99.5\%, sonar in 99.0\%, grok-4.5 in 87.0\%, and
mistral-large in 27.7\%.

\textbf{The number 42 splits the field.} gemini-3.7-flash, grok-4.5 and mistral-large rank
it fifth of 49 (7.5\%, 8.3\% and 7.2\% of returned numbers) and claude-sonnet-5 eighth
(4.8\%); sonar ranks it 30th (0.2\%), and gpt-5.6-luna never returned it in 1{,}200
numbers. In the two archived samples 42 was drawn once each and ranked 48th and 49th of
49, which is what chance does to some number at these sample sizes; neither ranking
supports a story about 42.

\textbf{Number ranges.} Model-level shares of numbers 1-31 ran 63.8\% to 68.0\% against
the uniform $31/49 = 63.3\%$; the two archived samples sat at 67.3\% and 68.8\%. The mean
returned number per engine ran 24.42 to 27.00 against a uniform expectation of 25.0.

\textbf{Data quality.} Of 1{,}200 attempted calls, 1{,}188 returned; all 12 failures are
HTTP 429 rate limits on mistral-large during the parallel run, a collection artifact. Four
claude-sonnet-5 answers (2.0\%) did not parse to six numbers and are excluded, with the
verbatim text kept in the raw file. No parsed answer contained a duplicate or an
out-of-range number. This leaves 1{,}184 valid tickets and 7{,}104 number occurrences.

\section{Related work}

That identical prompts yield different answers under nominally deterministic settings is
established \cite{atil2024}. Biased generated numbers are also documented: language models
produce biased binary sequences when asked to flip coins \cite{vankoevering2024} and
biased numbers across ranges, temperatures and prompt languages \cite{coronado2025}; both
lines score models against a theoretical target distribution. Earlier work decomposes
the variance of repeated brand answers \cite{zatuchin2026noise} and formalizes
repeated-query audit protocols \cite{zatuchin2026dice}. New here are the six-of-49 task,
the concentration of complete unordered tickets, and a physical-lottery comparison at
matched small sample sizes. The identical modal ticket in two systems is a descriptive
observation.

\section{Scope and limitations}

This study rejects independent uniform six-of-49 sampling under the tested configurations;
a biased stochastic process is still random. Output variation alone does not establish
uniform sampling or broad coverage of the intended answer space. The engines' outputs were
concentrated on a small set of recurring tickets, and the experiment does not distinguish
memorisation, decoding behaviour, prompt conditioning, retrieval, request routing, or
response caching as the cause. In particular, we did not verify that repeated calls
bypassed provider-side response caches; a cache returning stored answers for repeated
prompts would concentrate outputs by construction. sonar is retrieval-enabled by default
while the other five configurations received no search parameters, so the six systems ran
under two tool configurations and no single shared mechanism should be inferred across
them.

The physical baseline comes from the operator's statistics page
(\url{https://www.lotto.pl/lotto/statystyki}), retrieved 1 September 2026. The windows are
the matched 1 June to 31 August periods of 1999 and 2026, chosen as two same-season
samples 27 years apart. Both windows are archived as per-number integer count vectors
(\texttt{lotto\_1999.txt}, \texttt{lotto\_2026.txt}) in the public pack, so every baseline
value can be recomputed from integer counts without the live page; no computation uses
displayed percentages. The archive holds marginal counts only, and draw-level ticket
composition is unavailable, so this note makes no claims about repeated whole physical
tickets; the ticket-level analysis covers the engines alone. The number 44 was never drawn
in the 1999 window, a true zero.

\section{Conclusion}

Across the six tested configurations, the modal unordered ticket accounted for 22.5-68.0\%
of valid responses. This concentration is incompatible with independent uniform selection
of six distinct numbers from 1-49. The two archived Lotto samples were compatible with the
sample-size-specific simulation benchmarks used here. For anyone sampling model output,
run-to-run variation is no evidence of coverage: every engine put between 22.5\% and
68.0\% of its answers on one ticket.

\textbf{Future directions.} (1) Other languages: the collector's Polish arm
(\texttt{pl\_base}) is built and unrun, and favourite numbers may be locale-specific.
(2) Temperature sweeps: how $N_{\mathrm{eff}}$ moves with decoding temperature.
(3) Fixed seeds: whether seeded sampling narrows the ticket set further.
(4) The same design on commercial questions, where the concentrated set is brands; our
companion note measures that repertoire \cite{zatuchin2026saturation}.
(5) The number-theoretic structure of the modal tickets: they step across the wheel in
near-even gaps, which invites a formal description.
(6) Ticket-level comparison with the human favourite-number literature.

\section*{Declarations}

\textbf{Competing interests.} The author is CEO of Rankfor.AI, which sells AI-visibility
measurement; this note is methodological and names no customer.
\textbf{Data and code.} All raw answers, per-number counts, metrics, the collector, and
the recomputation script are public at
\url{https://github.com/Rankfor/rankfor-open/tree/main/research/lotto-models}.
\textbf{Provenance.} The Lotto baseline is the operator's published draw statistics
(\url{https://www.lotto.pl/lotto/statystyki}) for 1 June to 31 August 1999 and 2026,
retrieved 1 September 2026 and archived as per-number count vectors in the public pack.

\appendix

\section{Prompt variants and per-variant results}\label{app:prompts}

Each engine received 50 calls per variant. The four prompts, verbatim:

\begin{quote}\small
\texttt{en\_base}: ``Give me six different random numbers between 1 and 49.\\
Answer with the six numbers only, separated by commas.''\\[3pt]
\texttt{en\_lottery}: ``Pick six lottery numbers between 1 and 49.''\\[3pt]
\texttt{en\_terse}: ``I need 6 random numbers from 1 to 49. Just the numbers.''\\[3pt]
\texttt{en\_integers}: ``Generate six unique random integers in the range 1-49.''
\end{quote}

\begin{table}[ht]
\centering
\caption{Per-variant results: 50 attempted calls per cell. $N_{\mathrm{eff}}$ is the
effective diversity of number frequencies within the cell; the modal share is over the
cell's valid answers.}\label{tab:variants}
\footnotesize
\setlength{\tabcolsep}{5pt}
\begin{tabular}{llrrrrr}
\toprule
engine & variant & attempted & valid & distinct tickets & modal share & $N_{\mathrm{eff}}$ \\
\midrule
claude-sonnet-5 & en\_base & 50 & 50 & 13 & 36.0\% & 10.8 \\
claude-sonnet-5 & en\_lottery & 50 & 50 & 17 & 32.0\% & 11.4 \\
claude-sonnet-5 & en\_terse & 50 & 50 & 16 & 26.0\% & 13.3 \\
claude-sonnet-5 & en\_integers & 50 & 46 & 13 & 28.3\% & 12.0 \\
gemini-3.7-flash & en\_base & 50 & 50 & 14 & 62.0\% & 9.0 \\
gemini-3.7-flash & en\_lottery & 50 & 50 & 10 & 64.0\% & 9.3 \\
gemini-3.7-flash & en\_terse & 50 & 50 & 12 & 46.0\% & 9.4 \\
gemini-3.7-flash & en\_integers & 50 & 50 & 14 & 48.0\% & 10.2 \\
gpt-5.6-luna & en\_base & 50 & 50 & 3 & 96.0\% & 6.2 \\
gpt-5.6-luna & en\_lottery & 50 & 50 & 2 & 96.0\% & 6.5 \\
gpt-5.6-luna & en\_terse & 50 & 50 & 3 & 58.0\% & 11.9 \\
gpt-5.6-luna & en\_integers & 50 & 50 & 5 & 44.0\% & 12.9 \\
grok-4.5 & en\_base & 50 & 50 & 36 & 8.0\% & 23.1 \\
grok-4.5 & en\_lottery & 50 & 50 & 9 & 72.0\% & 7.4 \\
grok-4.5 & en\_terse & 50 & 50 & 30 & 28.0\% & 18.4 \\
grok-4.5 & en\_integers & 50 & 50 & 31 & 10.0\% & 21.1 \\
mistral-large & en\_base & 50 & 45 & 6 & 64.4\% & 7.7 \\
mistral-large & en\_lottery & 50 & 50 & 2 & 66.0\% & 7.1 \\
mistral-large & en\_terse & 50 & 48 & 13 & 37.5\% & 11.9 \\
mistral-large & en\_integers & 50 & 45 & 8 & 60.0\% & 9.1 \\
sonar & en\_base & 50 & 50 & 9 & 80.0\% & 8.9 \\
sonar & en\_lottery & 50 & 50 & 9 & 68.0\% & 10.2 \\
sonar & en\_terse & 50 & 50 & 8 & 82.0\% & 8.0 \\
sonar & en\_integers & 50 & 50 & 12 & 42.0\% & 12.7 \\
\bottomrule
\end{tabular}
\end{table}

\clearpage
\section{Deployment record}\label{app:deploy}

\begin{table}[ht]
\centering
\caption{Deployment record per engine. All calls ran on 1 September 2026, 09:05 to 09:08
UTC, one user message per call with a fresh context. No temperature, seed, or top-p
parameters were sent; provider-side defaults for decoding are undisclosed by the endpoints
and recorded as unknown. No search parameters were sent to any endpoint; sonar performs
retrieval by default. The returned model id is the id string in each API response, uniform
across all 200 calls per engine.}\label{tab:deploy}
\footnotesize
\setlength{\tabcolsep}{3.5pt}
\begin{tabular}{lllll}
\toprule
engine & requested model id & returned model id & endpoint host & extra parameters \\
\midrule
gpt-5.6-luna & gpt-5.6-luna & gpt-5.6-luna & api.openai.com & none \\
claude-sonnet-5 & claude-sonnet-5 & claude-sonnet-5 & api.anthropic.com & max\_tokens=128 \\
gemini-3.7-flash & gemini-3.7-flash & gemini-3.7-flash & generativelanguage.googleapis.com & none \\
grok-4.5 & grok-4.5 & grok-4.5 & api.x.ai & none \\
mistral-large & mistral-large-latest & mistral-large-latest & api.mistral.ai & none \\
sonar & sonar & sonar & api.perplexity.ai & none \\
\bottomrule
\end{tabular}
\end{table}

\end{document}